\documentclass[]{JHEP3}

\usepackage{graphicx} \usepackage{dcolumn}
\usepackage{bm}
\newcommand{\be}{\begin{equation}} \newcommand{\ee}{\end{equation}}

\newcommand{\bea}{\begin{eqnarray}} \newcommand{\eea}{\end{eqnarray}}

 \def\simgt{\stackrel{>}{{}_\sim}}

\title{de Sitter vacua from uplifting D-terms in effective
supergravities from realistic strings}
\author{A.~Ach\'ucarro \\ Lorentz Institute of Theoretical Physics,
Leiden University \\ 2333 RA Leiden, The Netherlands \\ E-mail:
\email{achucar@lorentz.leidenuniv.nl} \\ {\rm and} \\  Department of
Theoretical Physics, University of the Basque Country  UPV-EHU \\
48080 Bilbao, Spain} \author{B.~de Carlos \\ Department of Physics and
Astronomy, University of Sussex, \\ Falmer, Brighton BN1 9QJ, UK \\
E-mail: \email{B.de-Carlos@sussex.ac.uk}} \author{J.A.~Casas \\
Instituto de F\'{\i}sica Te\'orica, C-XVI, UAM \\ 28049 Madrid, Spain
\\ E-mail: \email{alberto.casas@uam.es} } \author{L.~Doplicher \\ Dipartimento
di Fisica "Enrico Fermi", Universit\`a di Pisa \\ Largo Pontecorvo 3,
Pisa, Italy \\ E-mail: \email{doplicher@df.unipi.it}}

\abstract{ We study the possibility of using the D-term associated to
an anomalous U(1) for the uplifting of AdS vacua (to dS or Minkowski
vacua) in effective supergravities arising from string theories,
particularly in the type IIB context put forward by Kachru, Kallosh, Linde and Trivedi (KKLT). We find a
gauge invariant formulation of such a scenario (avoiding previous
inconsistencies), where the anomalous D-term cannot be cancelled, thus
triggering the uplifting of the vacua. Then, we examine the general
conditions for this to happen. 
Finally, we illustrate the results by presenting different successful
examples in the type IIB context.}


\preprint{IFT-UAM-CSIC-05-52}

\begin{document}

\section{Introduction}

Our increasingly precise knowledge of the evolution of the universe,
condensed in the  $\Lambda$CDM model, has confirmed inflation as a
cornerstone of modern cosmology. This was an epoch of accelerated,
deSitter (dS) expansion in the distant past. There is also strong

observational evidence that the universe is entering a second phase of
accelerated expansion today, suggesting a positive vacuum energy
at present~\cite{WMAP}.

It would be desirable that this progress in the area of cosmology were
matched by analogous developments on the Particle Physics side. The
outstanding problem there is the development of a consistent quantum
theory of the fundamental interactions that would explain cosmological
observations. It is widely believed that string theory
fulfils all necessary requirements to become such fundamental theory,
but there are a number of technical issues that are hampering
progress. In particular, it is very difficult to find stable dS vacua,
which is directly related to the problem of moduli stabilisation.
  
Recently, we have witnessed outstanding progress in this area, perhaps
best represented by the results of Kachru et al.~\cite{Kachru:2003aw},
known from now on as KKLT. They devised a method for constructing dS
vacua through a combination of D-branes, fluxes and non-perturbative
effects, which we will explain in detail in the next section, as a
preamble to our work. The idea was, in the context of Type IIB theory,
to stabilise most of the moduli by fluxes at a high scale, and to
write down an effective low-energy theory for the remaining, so far
flat, $T$-modulus. These would be stabilised by  adding non-perturbative
effects to the superpotential.  The resulting vacuum happens to be
purely supersymmetric (SUSY) and anti deSitter (AdS). The idea, then, is
to add the effect of anti D3-branes to the model, which would act as 
an extra piece in the potential, breaking SUSY
explicitly and uplifting the vacuum to dS.

There are  a number of issues raised about this mechanism and its
successful implementation within a realistic string model.  In
particular, the explicit supersymmetry breaking introduced by the anti
D3-brane makes it difficult to compute corrections reliably, so  it
would be desirable to obtain dS vacua without explicit SUSY
breaking. A suggestion by Burgess et al.~\cite{Burgess:2003ic}, BKQ
from now on, consisted of dropping the anti-D3 branes and considering
D-terms coming from an anomalous U(1) instead. In favourable
conditions the D-terms could act as an uplifting potential.  This
approach was subsequently criticised by Choi et al.~\cite{Choi:2005ge}
and de Alwis~\cite{deAlwis:2005tf}. Quoting old results in
Supergravity, a model with vanishing F-terms must have also 
vanishing D-terms. Therefore
it is imperative that, in order for the D-term to act as an
uplifting potential, the F-terms have to necessarily break SUSY
(which was not the case for BKQ).

It is this whole issue that we want to address in the present
letter. We will reconcile the results of BKQ, on the one hand,
and Choi and de Alwis on the other, by exploiting the need for the
superpotential to be gauge invariant once an anomalous U(1) is
considered in the model.  We will show that a gauge invariant version of
BKQ can lead to stable dS vacua with realistic values of the
parameters in specific examples of Type IIB compactifications.

The paper is organised as follows: in 
section~\ref{sec:back}, we give a comprehensive review of the
structure of the KKLT mechanism and the subsequent proposal of BKQ. In
sections~\ref{sec:gaugeinv} and \ref{sec:implic} we look at the
consequences of implementing gauge invariance in a consistent way. In
section~\ref{sec:anomcan} we give a summary of the anomaly
cancellation conditions one has to take into account to work out
consistent examples.  In section~\ref{sec:examples} we present viable
examples in the context of Type IIB theory and, in
section~\ref{sec:conclu}, we conclude.

\section{Background}
\label{sec:back}

As already mentioned in the introduction, KKLT found an explicit way
to construct 4D de Sitter solutions of string theory. 
Their proposal was to use background fluxes for both NS and RR forms
to fix the complex-structure  moduli of a Calabi-Yau compactification
in the context of Type IIB theory.  They considered just one overall
K\"ahler modulus, $T$, not fixed by the fluxes,   which they stabilise
through non-perturbative effects.   In the language of  N=1
Supergravity (SUGRA), this is equivalent to considering a model with
K\"ahler potential
\be  K = -3\log (T+\bar{T}) \;\;,
\label{KIIB1}
\ee
and superpotential
\be  W = W_0 + A {\rm e}^{-a T} \;\;.
\label{WIIB}
\ee
$W_0$ is an effective parameter, coming from having integrated out all
complex-structure moduli through the use of fluxes~\cite{fluxes}, and $A,
a$ are constants.  The non-perturbative superpotential is either
generated by Euclidean D3-branes or by gaugino 
condensation~\cite{Derendinger:1985kk,Dine:1985rz,Burgess:1995aa} in a
non-abelian sector of $N$ wrapped D7-branes.  Notice here that the
gauge kinetic functions in the effective SUGRA theory are given (up to
${\cal O}(1)$ factors, see sect.~5 below) by
\be  
f_a = {T\over 2\pi}\;\;.
\label{fIIB1}
\ee
The corresponding scalar potential [given just by the F-part, $V_F$, see eqs~(\ref{VFVD}) below] has a
SUSY-preserving,  AdS minimum for $T$, which thus gets stabilised, but
at a negative energy.  In order to promote these AdS vacua to dS
vacua,  the authors add anti D3-branes to the construction, which they
quantify  in terms of a new piece in the potential, namely
\be  V = V_F + \frac{k}{T_R^2} \;\;,
\label{upKKLT}
\ee
where $k$ is a constant and $T_R
= {\rm Re}\;T$. If $k$ acquires a suitable value, then this term can
uplift the AdS vacuum obtained just with $V_F$. Notice that the
$k/T_R^2$ piece breaks supersymmetry explicitly which, besides
aesthetic reasons, complicates the analysis and reduces our control
over the effective theory.

In ref.~\cite{Burgess:2003ic} BKQ proposed an attractive variation of
KKLT to avoid such shortcomings. Namely, instead of  anti D3-branes,
they considered  the possibility of turning on fluxes for the gauge
fields living  on D7-branes. These fluxes would, in turn, generate a
Fayet-Iliopoulos (FI)  term in the  4D effective action of the form
\bea
\label{VDFernando}
V_D  = {1\over 2 g_{\rm YM}^2}D^2 = {\pi \over T_R}\left({E\over T_R}
+  \sum_I q_i |\varphi_i|^2\right)^2 \;\;, \eea
where $T_R=2\pi/g^2_{\rm YM}$, $E$ is some constant arising from
non-trivial fluxes for the gauge fields living on the D7-branes,
$\varphi_i$ are the scalar components of the chiral superfields,
except $T$, and $q_i$ are the corresponding charges under the
anomalous U(1)$_X$ group. The previous equation assumes a minimal
K\"ahler potential for $\varphi_i$, which is clearly a simplification,
but it does not affect crucially the results.

The next assumption of BKQ is that (at the minimum) all
$\langle\varphi_i\rangle=0$. They argue that this can arise in two
different ways: if the $q_i$ charges are such that the above D-term
cannot be cancelled, it is reasonable to assume that the
$\varphi_i$ fields will settle at the origin to minimise $V$.  As the
authors admit, this ``non-cancellability'' is a crucial assumption
which, so far,  has not proved to be realised in
practice. Alternatively, if the superpotential does not depend on the
$\varphi_i$ fields [as happens in eq.~(\ref{WIIB})], then the
corresponding  F-term contributions to the potential, $|D_iW|^2\sim
|\varphi_i W|^2$ may be efficient to render  $\varphi_i=0$ at the
minimum. In either case, they can replace
\bea
\label{VDFernando2}
V_D\sim {\pi E^2 \over T_R^3}\ , 
\eea
which has a form similar to the non-SUSY potential suggested by KKLT
in (\ref{upKKLT}) and, indeed, can work as an uplifting term: by
adding the piece (\ref{VDFernando2}) with an appropriate value of $E$
to the $V_F$ potential,  the initial KKLT AdS minimum becomes a dS
vacuum, as desired.

\vspace{0.3cm} The results of BKQ have been criticised by Choi et al.~\cite{Choi:2005ge} 
(and also in ref.~\cite{deAlwis:2005tf}). Their
argument is based on the simple relation between the value of the
D-terms and the F-terms in a (D=4, N=1) SUGRA theory, namely
\bea
\label{formula}
D^a ={i\over {\rm Re}(f_a)}{1\over W} \eta^I_a D_I W \;\;, \eea
where $\Phi^I$ are the chiral superfields, $\eta^I_a(\Phi)$ denote
their gauge transformation, $\delta \Phi^I= \epsilon_a \eta^I_a(\Phi)$
($\epsilon_a$ are the infinitesimal parameters of the
transformation), $D^a = i\eta^I_a K_I/{\rm Re}(f_a)$, and $D_I$
is the K\"ahler derivative: $D_I W =  \partial_I W + (\partial_I K)W$.
($X_I\equiv \partial_I X = 
\partial X/\partial \Phi_I$, as usual).

Eq.~(\ref{formula}) is valid at any point in field space (i.e. not
necessarily at the minimum of the potential), except where  $W=0$, and
can be easily proved using the gauge invariance of $W$ and the general
form of the F-part and the D-part of the scalar potential, given by
\bea
\label{VFVD}
V_F &=& e^K\left( K^{I\bar J}D_IW D_{\bar J} \bar{W} -3|W|^2\right)
\;\;, \nonumber\\ V_D  &=& {1\over 2}{\rm Re}(f_a) D^aD^a =  {1\over 2
{\rm Re}(f_a)}\left(i\eta^I_aK_I\right)^2 \;\;.  \eea
($K^{I\bar J}$ is the inverse of the  $K_{I\bar J}=\partial^2 K
/\partial \Phi_I\partial \bar \Phi_J$ matrix.)
As mentioned above, when {\em just} the F-part of the KKLT potential
(which coincides with the BKQ one)  is minimised, the AdS
SUSY-preserving minimum is at $D_T W=0$, $W \neq 0$.  If one then
plugs $V_D$ in, as done by BKQ,  eq.~(\ref{formula}) guarantees that
that point in field space is still stationary with $D=V_D=0$.  Hence
adding $V_D$ does not lift the AdS minimum.

More generally, it is clear that, if the set of equations  $D_I W =0$
can be solved simultaneously, this corresponds to an AdS (if $W\neq
0$) SUSY minimum. Then  the previous  argument applies and $V_D$ does
not act as an uplifting potential~\footnote{However, even in that case there could still  be local dS
minima, which might be good enough for phenomenological purposes; this
might also work just without the aid of any $V_D$ potential or
similar.}.

The results of BKQ and the argument of Choi et al. seem  contradictory
in appearance: if Choi et al. are right, BKQ should not have found an
uplifted dS minimum once the $V_D$ piece is added.  This contradiction
is solved once gauge invariance is properly implemented, as we discuss
in the next section.

\section{Implementation of gauge invariance and its consequences}
\label{sec:gaugeinv}

From the general form of $V_D$ in (\ref{VFVD}), it is clear that a FI
term such as that of eq.~(\ref{VDFernando}) implies 
that the $T$ modulus is charged under U$(1)_X$, that is, $\eta^T_X\neq 0$.
More precisely, from eqs~(\ref{VDFernando}) and (\ref{VFVD}), we can
write the auxiliary $D_X$ field as
\bea
\label{DXFernando}
D_X  = -g_X^2\left({E\over T_R} + \sum_I q_i |\varphi_i|^2\right) =
ig_X^2\eta^I_aK_I\;\;, \eea
implying
\bea
\label{etaT}
\eta^T_X\equiv i{\delta_{GS}\over 2}={-2iE\over 3}\;.  \eea
This indicates that the non-perturbative superpotential term
$Ae^{-aT}$  in (\ref{WIIB}) is not U(1)$_X$ gauge
invariant~\footnote{Actually, BKQ mentioned this point in their
article but, for simplicity, they did not  attempt to incorporate gauge
invariance in the analysis.}, which is the reason why the BKQ
potential does not obey the Choi et al. argument~\footnote{If the U(1)
symmetry is a (gauged) R-symmetry, the non-perturbative superpotential
can transform with the correct phase under a U(1) transformation. Then
the FI term is purely constant~\cite{Binetruy:2004hh}. Notice,  however, that the constant $W_0$ in
eq.~(\ref{WIIB}) does not transform (as it should). The  possibility
of using a U(1) R-symmetry of this kind, with $W_0=0$, to get dS
vacua, thanks to the interplay between $V_F$ and $V_D$, has been
explored in ref.~\cite{Villadoro:2005yq}.}.  The
implementation of gauge invariance turns out to be crucial, not only
for the consistency of the approach, but also for the qualitative
features of the results. We will see also that many interesting
characteristics of the potential and its minima can be predicted just
on grounds of gauge invariance.

The analysis of this section and the next one,  although described in
terms of the Type IIB set-up used by KKLT and BKQ, is valid for any
SUGRA model, in particular for those coming from heterotic string
constructions.

The first step is to write the K\"ahler potential and the
superpotential in a gauge-invariant form. For the K\"ahler potential,
$K$, we follow the usual prescription
\be  K = -3\log (T+\bar{T})\;\;\;  \longrightarrow\;\;\;   K = -3\log
(T+\bar{T}+\delta_{GS}V)\;\;,
\label{KIIBgauge}
\ee
where $i{\delta_{GS}\over 2}\equiv\eta^T_X$ and $V$ is the U(1)$_X$
vector superfield, while the non-perturbative superpotential  should
be written as
\bea
\label{WNP}
W_{np}=Ae^{-aT}\;\;\; \longrightarrow\;\;\; W_{np}=J(\varphi_i)e^{-aT}
\;\;, \eea
where $J(\varphi_i)$ is an analytic function transforming  under
U(1)$_X$ just opposite  to $e^{-aT}$, i.e. $J(\varphi_i)\rightarrow
e^{ia\delta_{GS}/2} J(\varphi_i)$, and invariant under the other gauge
groups~\footnote{ Incidentally, notice that an exponential for the
$T$-superpotential is essentially the only functional form that can be
made gauge invariant by the action of the matter superfields (which
transform as a phase under U(1)$_X$), which is a remarkable fact.
Note also that for $W_0 \neq 0$ the only option consistent with gauge
invariance is is to make $W$ and $K$ separately gauge invariant 
(rather than invariant up to K\"ahler transformations).}.

The first conclusion is that one of the alternative assumptions of BKQ
to get $\langle\varphi_i\rangle=0$, namely to suppose  $W\neq
W(\varphi_i)$, cannot occur in practice.  Next, we analyse the other
alternative, i.e. the possible non cancellability of the FI potential
due to the signs of the U(1)$_X$ charges, which  is an interesting
problem on its own.

A sufficient condition for the cancellability of the D-terms in a
supersymmetric theory~\cite{Buccella:1982nx,Gatto:1986bt,Font:1988mm}
is the existence of an analytic function, $\tilde J(\varphi_i)$,
invariant under all the gauge groups except the anomalous U(1)$_X$,
and having an anomalous charge with sign opposite to the
$\varphi_i$-independent term in $D_X$, i.e. $E/T_R$ in
eq.~(\ref{DXFernando}).  Now, from the above transformation properties
of $J(\varphi_i)$ (assuming positive $a$, as usual) and
eq.~(\ref{etaT}), it is clear that $J(\varphi_i)$ precisely fulfils
the previous requirements, i.e.  we can identify
$J(\varphi_i)\equiv\tilde J(\varphi_i)$ in eq.~(\ref{WNP}). This
means that there is a set of $\langle \varphi_i\rangle$ values that
cancel all the D-terms, including the anomalous one.  In other words,
the BKQ assumption that the anomalous FI term cannot be cancelled is
not consistent with the presence of a non-perturbative {\it
analytic} superpotential.

The apparent conclusion from the previous paragraphs is that the BKQ
proposal of using FI terms to uplift the potential is definitely
hopeless.  However, things are more subtle in practice. $W_{np}$ is
typically an effective superpotential, so its analyticity does not
need to be guaranteed in the whole range of the $\varphi_i$
fields. Actually, this is the case when $W_{np}$ is originated by
gaugino condensation.  For example,  for SU($N$) with $N_f$
``quark'' pairs, $\{Q_j, \bar Q_j\}$, $W_{np}^{(eff)}$ is as in
eq.~(\ref{WNP}) with~\cite{Affleck:1983mk,Taylor:1982bp}
\bea
\label{JSUN}
J(\varphi)=(N-N_f) (\det M^2)^{-1\over N-N_f} \;\;, \eea
where $(M^2)^i_j=2Q^i\bar Q_j$.  The important point is that the
exponent of $\det M^2$ in eq.~(\ref{JSUN}) is {\em negative} (at least
for $N_f<N$). Therefore, $J(\varphi)$ cannot play the role of the
{\em analytic} function $\tilde J(\varphi)$ in the above condition for
the cancellation of the D-terms.  $J(\varphi)^{-1}$ (or, more
precisely, $J(\varphi)^{-(N-N_f)}\sim  \det (Q\bar Q)$) might do it,
but it has the wrong sign for the cancellation of $D_X$. In
consequence, with this $W_{np}^{(eff)}$ it is possible, in principle,
that the $D_X$ term cannot be cancelled by any choice of
$\langle\varphi_i\rangle$. This rescues the BKQ assumption.

Nevertheless, even in this case, the fact that $W_{np}$ depends
crucially on $\varphi_i$  alters significantly the BKQ analysis. In
particular, $\langle\varphi_i\rangle$ cannot be simply approximated by
zero (which presents a singularity) or replaced by a constants in
$W_{np}$. As will be clear in sect.~6, these fields
contribute their own (sizeable) part to the effective
potential and  cannot be ignored for minimisation issues.
Consequently, the BKQ analysis must be
redone incorporating the $\varphi_i$ fields. We leave this task for
section~\ref{sec:examples} and discuss next other relevant
implications of gauge invariance.

\section{Further implications for gaugino condensation and FI terms}
\label{sec:implic}

We have just seen that a non-perturbative superpotential produced by
gaugino condensation may be compatible with a non-cancellable
anomalous FI term, thus rescuing the BKQ hypothesis.  It is
interesting, however, to point out  the circumstances under which 
this {\em cannot} happen.

First of all, it is quite obvious that, if the condensing group has no
matter representations, the condensing superpotential is not only
incompatible with a non-cancellable FI term, but with the existence of
a FI term at all~\footnote{Blanco-Pillado et al.~\cite{Blanco-Pillado:2005xx} 
have argued that, in the presence of D-
anti D-brane pairs, the four dimensional effective action should also
include a constant FI term coming from the tension of the branes. Here
we do not consider such a possibility.}.  This occurs
because the presence of the FI term indicates a non-trivial
transformation of the $T$-field, which can only be compensated in the
non-perturbative superpotential (\ref{WNP}) by the transformation of
$J(\varphi)$, which is a function of the matter representations, as
given in eq.~(\ref{JSUN}).

Second, if the fields transforming under the condensing group are
massive, gaugino condensation will be again incompatible with a
non-cancellable FI term. By `massive' we mean that the superpotential
contains operators with the form of a mass term, with (generically)
field-dependent masses. For example, ordinary Yukawa operators are
mass terms from this point of view.  Since in the gaugino condensation
literature the massive case is very common, it is appropriate to
discuss this point more in depth. Working again with SU($N$) with
$N_f$  $\{Q, \bar Q\}$ pairs, if these fields are massive the
superpotential contains a term
\bea
\label{Wmass}
W\supset [{\cal M}(A)]^{ij}Q_i \bar Q_j \;\;, \eea
where $A$ represents other chiral fields (with possible VEVs, but this
is not relevant in the following discussion). The gauge invariance of
$W$ implies that $\det[{\cal M}(A)]$ transforms under U(1)$_X$ exactly
opposite to $\det [Q\bar Q]\propto \det M^2$.  Therefore we can simply
use $\det[{\cal M}(A)]$ as the $\tilde J(\varphi)$ function of the
cancellation condition [its charge has the correct sign under
U(1)$_X$]. This guarantees the existence of $\langle A\rangle$ values
cancelling the FI D-term.

This statement is confirmed by the fact that, in the massive case, the
chiral superfields  can be integrated out, so that $W_{np}^{(eff)}$
reads~\cite{Taylor:1982bp,Lust:1990zi,deCarlos:1991gq}
\bea
\label{Wmass2}
W_{np}^{(eff)}\propto [\det{\cal M}(A)]^{1\over N} e^{-a T} \;\;. \eea
This has the form (\ref{WNP}), with  $J(\varphi)$ similar to
eq.~(\ref{JSUN}), but now with a positive exponent.  Therefore we can
use $\det {\cal M}(A)$ as the  analytic $\tilde J(\varphi)$ function
of the D-cancellation condition~\footnote{It has been argued
elsewhere~\cite{deCarlos:1991gq}  that the general form of
$W_{np}^{(eff)}$ for a generic condensing group (not necessarily
SU($N$) with $N_f$ flavours) is $W_{np}^{(eff)}\propto
[m(A)]^{3(1-\tilde\beta/\beta)}e^{-3/2g^2\beta}$, where $\beta$
($\tilde\beta$) is the $\beta$-function of the gauge group with $N_f$
($N_f=0$) flavours.  Again the exponent is positive, guaranteeing the
cancellation of the D-term.}.

On the other hand, many authors  have argued that the limit
$m\rightarrow 0$ is singular~\cite{Taylor:1982bp}, or it leads to the
dissappearance of an effective supersymmetric
superpotential~\cite{Affleck:1983mk}. According to that,  it could
seem that the only acceptable case is the massive case and, in
consequence, we would be back  to the strong statement that the D-term
must be neccesarily cancelled by some choice of
$\langle\varphi_i\rangle$.  However, as we discuss next, those
arguments can be revisited to see that the formulation with massless
matter is not problematic in this case.

Based on arguments related to the Witten index for global
SUSY~\cite{Witten:1981nf,Weinberg:2000cr}, the authors of the previous 
references have
argued that the $m\rightarrow 0$ case cannot be described by a 
non-trivial superpotential with SUSY preserving vacuum.
In particular, for a large class of global SUSY scenarios,
Affleck et al.~\cite{Affleck:1983mk} have
shown that, for $m\rightarrow 0$, the minimisation of
the potential leads to $D_M W\rightarrow 0$, and hence to
$M^2\rightarrow \infty$ (run-away behaviour)
and $W_{np}\rightarrow 0$. This indicates that there
is no effective SUSY superpotential describing the condensation of
gauginos in that case, which is consistent with the mentioned
general expectations.

In our case we have some differences with the previous scenarios:
a Supergravity framework, the presence of the constant
flux piece, $W_0$, and the dependence on the $T$-field. Although we see the
mentioned behaviour in the global SUSY limit, in the complete
case things are different.  Now we have two conditions for preserving SUSY, $D_M W=0$
and $D_T W=0$. If both
could be satisfied at the same time, then we would have a SUSY
minimum. [Notice that this would not be changed by the inclusion of
the FI term since, from eq.~(\ref{formula}), this would cancel at that
point.] Then, we would violate the Witten index arguments, at least
for a minimum with zero vacumm energy (notice that, for unbroken SUSY, 
this requires $\langle W\rangle=\langle \partial_I W\rangle=0$,
which includes the condition for unbroken SUSY in the global limit). However,
if the two conditions {\em cannot} be satisfied at the same time, then
SUSY would be necessarily broken, recovering the agreement with the general
arguments. So we can have a consistent, non-singular, massless situation, 
provided SUSY is broken
(by $D_M W\neq0$ and/or $D_T W\neq0$).

It is remarkable that the same conclusion can be reached  using our
arguments based on the cancellability of the D-terms: if we are in a
theory with a non-cancellable D-term (which, as argued, is only
compatible with gaugino condensation in  the massless matter case),
then the equations $D_I W=0$  ($I$ running over all the chiral
superfields) and $W\neq 0$ cannot be fulfilled simultaneously.
Otherwise we would have a paradox since, at that point, $V_D$ should
also be zero because of eq.~(\ref{formula}), but this is not possible
by hypothesis. Therefore the only way out is that either the set of
equations $D_I W=0$ (i.e. $D_M W=0$ and $D_T W=0$ in the simplest
set-up)  has no solution, which is the conclusion reached in the
previous paragraph; or $W\rightarrow 0$, indicating a run-away
behaviour and, therefore, the disappearance of the non-perturbative
superpotential which describes the  gaugino condensation.

In section~\ref{sec:examples}, we will show explicitly with
examples how these results take place in practice.

\section{Aspects of  Type IIB and the Heterotic cases}
\label{sec:anomcan}

\subsection*{Type IIB}

As mentioned above, a relevant ingredient of the (Type IIB) set-up of
KKLT is the presence of a SU($N$) condensing group arising from stacks
of $N$ D7-branes wrapped on some 4-cycle of the Calabi-Yau space. It
should be noticed here that for each SU($N$) there typically appears a U(1)
factor [although this is not a strict rule].  Some of these U(1)s, or
combinations of them, can be anomalous.

Although the discussion is easier in terms of an overall (K\"ahler)
modulus, $T$, in general there will be several relevant moduli, $T_i$,
corresponding to the  independent 4-cycles.  The gauge kinetic
functions of the different gauge groups, $f_a$, are combinations of
these moduli, with coefficients that depend on the geometric structure
of the associated 4-cycle.  In other words, the kinetic terms for the
gauge superfields are of the form
\bea
\label{WW}
{\cal L}\supset \int d^2\theta {1\over 8\pi}\sum_i k_a^{(i)} T_i
W^\alpha_a W_{a\alpha} \;\;, \eea
where $W^\alpha_a$ is the field strength superfield of the $G_a$ gauge
group and  $k_a^{(i)}$ are positive, ${\cal O}(1)$ model-dependent 
constants~\cite{Ibanez:1998rf}.
Note that
eq.~(\ref{WW}) includes axionic-like couplings $\sim k_a^{(i)}({\rm
Im} \; T_i)\ F_a \tilde F_a$. Likewise, there may be couplings of the
${\rm Im} \;T_i$ to $R\tilde R$ ($R$ denoting the 4D Riemann tensor)
which depend on the particular geometrical structure of the model. 
As usual, if the $T_i$-fields transform non-trivially,
$T_i\rightarrow T_i + i{\delta^{(i)}_{GS}\over 2}\epsilon$, under a
particular U(1), say U(1)$_X$, this will be reflected in the presence
of $T_i$-dependent FI terms.  The corresponding  transformation of the
Lagrangian (\ref{WW}) is  $\propto  k_a^{(i)}\delta^{(i)}_{GS} F_a
\tilde F_a$, which has to be compensated, if different from zero,  by
the transformation arising from the $[G_a]^2\times {\rm U(1)}_X$
anomaly. This is the Green-Schwarz mechanism.  More precisely, for
$G_a\neq {\rm U(1)}_X$ this requires
\bea
\label{anom1}
\sum_i  k_a^{(i)}\delta^{(i)}_{GS} =-{1 \over \pi} \sum_r K(r) q_X(r)
\;\;,  
\eea
where $r$ runs over all the chiral superfields transforming under
$G_a$, and $K(r)$, $q_X(r)$ are the corresponding representation index
and U(1)$_X$ charge, respectively. For $G_a= {\rm U}(1)_X$ there is an extra $(1/3)$
factor
\bea
\label{anom2}
\sum_i  k_X^{(i)}\delta^{(i)}_{GS} =-{1 \over 3\pi} \sum q_X^3 \;\;,
\eea
where the second sum runs over all the states. If $G_a$ is a different
U(1), say U(1)$_a$, the $(1/3)$ factor disappears and $q_X^3$ is
replaced by $q_a^2q_X$. Similarly, the mixed gauge-gravitational
anomaly, proportional to $\sum q_X$ gets cancelled by the
(model-dependent) couplings of the moduli to $R\tilde R$. 
All the
remaining gauge anomalies (e.g. U$(1)_1\times {\rm U}(1)_2\times {\rm U}(1)_X$)
are vanishing.

If there is just one overall-modulus, $T$, the gauge group is
typically ${\rm SU}(N)\times {\rm U(1)}$. For several $T_i$, there can
be many ${\rm SU}(N)\times {\rm U(1)}$ factors, with different values
of  $k_a^{(i)}$, $\delta^{(i)}_{GS}$.
The Standard Model gauge group could arise in this way from  wrapped
D7-branes or, alternatively, from stacks of D3-branes
sitting at a point of the compactified space.  In the latter case the
gauge coupling is given by the dilaton $S$, instead of $T$.

In summary, although conditions (\ref{anom1}, \ref{anom2}) for anomaly
cancellation must be fulfilled, and they are crucial for the
consistency of the approach, there is a lot of freedom
(i.e. model-dependence) for the possible values of the charges and the
$k^{(i)}$ coefficients.

Let us give, for future use, the relevant formulae for the   overall
modulus case and an ${\rm SU}(N)\times {\rm U(1)_X}$ gauge group with
$N_f$ quark pairs, $\{Q_j, \bar Q_j\}$, with charges $q$ and $\bar q$
respectively. For simplicity we will consider also an overall squark
condensate $|M|^2\equiv |M_1|^2=|M_2|^2=\ldots$, where $M_i=\sqrt{2
Q^i \bar{Q}_{\bar{i}}}$, with $i=1,\ldots,N_f$.  We are assuming here
$|Q^i|^2=|\bar{Q}_{\bar{i}}|^2\equiv |Q|^2$, which guarantees the
cancellation of the SU($N$) D-term.

The K\"ahler potential and the gauge kinetic functions for this system
are given (in $M_p$ units) by
\be
\label{KIIB}
K = -3 {\rm log}(T+\bar{T})  + \sum_{i=1}^{N_f} \left(|Q_i|^2 +
|\bar{Q}_{\bar{i}}|^2\right) =  -3 {\rm log}(T+\bar{T}) + N_f
|M|^2\;\;, \ee
\be
\label{fIIB}
f_N =  {k_N \over 2\pi} T,\;\;\;\;\;\;f_X =  {k_X \over 2\pi} T \;\;,
\ee
where $k_N$, $k_X$ are ${\cal O}(1)$ positive constants. 
The previous equation assumes a minimal
K\"ahler potential for $Q_i$, $\bar Q_i$, which is a simplification,
but not important for the present discussion, nor for the results
of the next section. Under a U$(1)_X$
transformation (with parameter $\epsilon$) $T$ transforms as
$T\rightarrow T + i{\delta_{GS}\over 2}\epsilon$.  The anomaly
cancellation conditions for ${\rm SU}(N)^2\times {\rm U}(1)_X$ and U$(1)_X^3$ 
read~\footnote{The sign of the right hand side of eq.~(\ref{anom3})
differs from the one quoted in~\cite{Binetruy:1996uv} 
(in the context of heterotic string effective Supergravity) since 
our convention for the quark charges is that, under a U$(1)_X$
transformation,
$Q\rightarrow {\rm e}^{{\rm i} q\epsilon}Q$,
$\bar Q\rightarrow {\rm e}^{{\rm i} \bar q\epsilon}\bar Q$,
while in their (implicit) conventions the sign of these exponents
was negative.}
\bea
\label{anom3}
\delta_{GS} =-{N_f (q + \bar q)\over 2 \pi k_N}  = -{N N_f (q^3 + \bar
q^3)\over 3 \pi k_X}   \;\;,  
\eea
where we have supposed for simplicity that, beside the quarks, there
are no other fields with non-vanishing U$(1)_X$ charge (otherwise,
the relations are straightforwardly modified). 
The effective condensation superpotential is given by
\bea W_{np} & = & (N-N_f)  \left(\frac{2
\Lambda^{3N-N_f}}{M^{2N_f}}\right)^{1\over N-N_f}  \nonumber\\ &=
&(N-N_f) \left(\frac{2}{M^{2N_f}}\right)^{1\over N-N_f}  {\rm e}^{-4
\pi k_N T\over N-N_f} \nonumber\\ &= &(N-N_f)
\left(\frac{2}{M^{2N_f}}\right)^{1\over N-N_f} {\rm
e}^{2N_f(q+\bar{q})T\over \delta_{GS}(N-N_f)} \;,
 \label{supncnf}
\eea
where in the last line we have used eq.~(\ref{anom3}). Notice that
$W_{np}$ has the form  (\ref{WNP}) and is indeed gauge invariant,  as
desired. In other words, the anomaly cancellation condition guarantees
the gauge invariance of $W_{np}$.
Incidentally, note that (contrary to recent claims~\cite{Villadoro:2005yq})
there is no problem in having a 'racetrack'
scenario~\cite{Krasnikov:1987jj,Dixon:1990ds,Casas:1990qi,deCarlos:1992da}, i.e. several condensation superpotentials with
different exponents, since all of them are rendered gauge-invariant
thanks to the presence of matter fields.
The D-part of the scalar potential is obtained from the
generic expression eq.~(\ref{VFVD})
\bea V_D &=& \frac{\pi}{4k_X T_R} \left(N_f(q+\bar{q})|M|^2  -
\frac{3\delta_{GS}}{2T_R}\right)^2 \nonumber\\ &=& \frac{\pi}{8k_N
T_R}{3(q+\bar q)\over N(q^3+\bar q^3)} \left(N_f(q+\bar{q})|M|^2  -
\frac{3\delta_{GS}}{2T_R}\right)^2\;\;.
\label{VDIIB}
\eea

\subsection*{Heterotic}

For the heterotic string case things are much more constrained. The
gauge kinetic functions are given by the dilaton field, $S=S_R + iS_I$, up to
Kac-Moody level factors, $k_a$ (positive integer-numbers, except for U(1)
groups) and one-loop string corrections,
\bea
\label{WWhet}
{\cal L}\supset \int d^2\theta {1\over 4}\sum_i k_a^{(i)} S W^\alpha_a
W_{a\alpha} \;\;.  
\eea
In consequence there can be only one anomalous $U(1)$, whose anomaly is 
cancelled by the transformation of $S$, which thus triggers
an FI term as explained above~\cite{Dine:1987xk}. 
For example, in conventional Calabi-Yau
and orbifold compactifications with $k_a=1$ (except for U$(1)$ groups), 
the  anomaly cancellation condition reads~\cite{Dine:1987xk,Casas:1987us}
\bea
-\delta_{GS} &=&{1 \over 2\pi^2} \sum_r K(r) q_X(r)
={1 \over 2\pi^2}{1\over 3k_X} \sum_n q_X^3=
{1 \over 2\pi^2}{1\over k_a} \sum_n q_X q_a^2=
{1 \over 2\pi^2}{1\over 24} \sum_n q_X\;\;, 
\nonumber \\
&&
\label{anom4} 
\eea
where the notation is as for eqs~(\ref{anom1}, \ref{anom2}), $q_a$
are the charges of the states under any U$(1)_a\neq {\rm U}(1)_X$ gauge group,
and $n$ runs over all the states.

The anomalous D-potential associated with U$(1)_X$ has the form
\bea V_D &=& \frac{1}{8k_X S_R} \left(\sum_n q_x^{(n)}K_i\phi_i
-\frac{\delta_{GS}}{4S_R}\right)^2 
\nonumber\\ 
&=& \left(\frac{\sum_n q_X}{\sum_n q_X^3}\right) {1\over S_R}
\left(\sum_n q_x^{(n)} K_n\phi_n
-\frac{\sum_n q_X}{192\pi^2}{1\over S_R}\right)^2 
\;\;,
\label{VDhet}
\eea
where $q_x^{(n)}$ is the anomalous charge of $\phi_n$ and
the explicit form of 
$K_n=\partial K/\partial \phi_n$ depends 
on the untwisted or twisted character, and
the corresponding modular weight, of the $\phi_n$-field, being in
general a function of the K\"ahler moduli \cite{Dixon:1989fj}.
(In the minimal K\"ahler simplification, $K_i=\phi^+$.)

The formulation of the gaugino condensation superpotential is
similar to the type IIB case.
For a ${\rm SU}(N)\times {\rm U(1)_X}$ gauge group with
$N_f$ quark pairs, $\{Q_j, \bar Q_j\}$, with charges $q$ and $\bar q$
respectively, and in the overall squark
condensate simplification, $|M_1|^2=|M_2|^2=\ldots\equiv |M|^2$
[$M_i=\sqrt{2
Q^i \bar{Q}_{\bar{i}}}$, with $i=1,\ldots,N_f$], 
the effective condensation superpotential is given by
\bea W_{np} & = & (N-N_f)  \left(\frac{2
\Lambda^{3N-N_f}}{M^{2N_f}}\right)^{1\over N-N_f}  
\nonumber\\ 
&=&(N-N_f) \left(\frac{2}{M^{2N_f}}\right)^{1\over N-N_f}  
{\rm e}^{-8\pi^2 k_N S\over N-N_f} 
\nonumber\\ 
&= &(N-N_f)
\left(\frac{2}{M^{2N_f}}\right)^{1\over N-N_f} {\rm
e}^{2N_f(q+\bar{q})S\over \delta_{GS}(N-N_f)} \;.
 \label{supncnfhet}
\eea
For the heterotic string, it is not possible to play freely with a
constant superpotential triggered by fluxes~\cite{Dine:1985rz}, so in order
to generate realistic non-trivial
minima for $S$ one needs more than one condensing groups. This is the
so-called 'racetrack mechanism'~\cite{Krasnikov:1987jj,Dixon:1990ds,Casas:1990qi,deCarlos:1992da} 
(for other
mechanisms to stabilise the dilaton see, for example,  
refs~\cite{Shenker:1990uf,Banks:1994sg,Casas:1996zi,Binetruy:1996xj,Barreiro:1997rp}). 
Then, suppose that the gauge group is SU$(N_1)\times {\rm SU}(N_2)\times {\rm U}(1)_X$.
For each condensate there is an effective superpotential of the form 
(\ref{supncnfhet}). Assuming that, beside the quarks, 
there are no other matter 
fields with non-vanishing $q_X$-charges, the anomaly cancellation conditions 
(\ref{anom4}) read now, with an obvious notation,
\bea
-\delta_{GS} &=&{1 \over 4\pi^2} N_{f1}(q_1 + \bar q_1)
={1 \over 4\pi^2} N_{f2}(q_2 + \bar q_2)
\nonumber\\
&=&{1 \over 6\pi^2 k_X} \sum_{j=1,2} N_j N_{fj}(q_j^3+\bar q_j^3)
={1 \over 48\pi^2} \sum_{j=1,2} N_j N_{fj}(q_j+\bar q_j)
\;\;.
\label{anom5} 
\eea
(If there are extra fields, the modification of these relations is 
straightforward.) Then $V_D$ has the form (\ref{VDhet}) with 
$\sum_n q_x = \sum_{j=1,2} N_j N_{fj}(q_j+\bar q_j)$,
$\sum_n q_x^3 = \sum_{j=1,2} N_j N_{fj}(q_j^3+\bar q_j^3)$.

It is worth-noticing that, in the absence of extra matter, the
anomaly cancellation condition forces $N=12$, independently
of the number of condensing groups. However, even in this case, a
racetrack is possible, since the number of flavours can be different
for each one.

\section{Examples in Type IIB}
\label{sec:examples}

We will now illustrate explicitly the results of the previous
sections,  in particular the realisation of uplifting D-terms in the
context of (N=1, D=4) effective supergravities from (type IIB) string
theory. The scenario can be considered as a simple, gauge invariant
version of the  BKQ model which cures previous inconsistencies and
allows for dS vacua.

\subsection{The structure of the scalar potential and SUSY breaking}

Let us consider the KKLT set-up with gaugino condensation as the
origin of the non-renormalisable superpotential. We will work in the
simplest case of a single overall modulus, $T$, and ${\rm SU}(N)\times
{\rm U(1)_X}$  gauge group with $N_f$ quark pairs, $\{Q_j, \bar
Q_j\}$, with anomalous charges $q$ and $\bar q$ respectively. We will also use
an overall condensate $|M|^2$, as discussed in the previous section.
There could be a singlet $\varphi$  as well, giving mass to the
condensate but, as argued in section~4, we are omitting it in order to
allow for a non-cancellable  D-term~\footnote{In
ref.~\cite{Dudas:2005vv} the issue of  gauge invariance was addressed
in the presence of massive matter  condensates. The resulting
formalism was applied to the derivation  of soft breaking terms, in
particular to estimating the contribution  of the resulting D-term.}.
The K\"ahler potential and the gauge  kinetic functions are thus given
by eqs.~(\ref{KIIB}, \ref{fIIB}), and the  superpotential reads
\be W=W_0 + W_{np}\;, \ee
where $W_0$ is the effective constant superpotential triggered by the
presence of fluxes and $W_{np}$ is the gauge-invariant gaugino
condensation superpotential, given by eq.~(\ref{supncnf})~\footnote{For a detailed study of the interplay between both terms, including the origi of $W_0$, see ref.~\cite{Derendinger:2006ed}.}. The scalar
potential $V=V_F+V_D$ can be explicitly written using the general
formulae (\ref{VFVD}). In particular, $V_D$ is given by
eq.~(\ref{VDIIB}).

Let us first show that this simple model breaks SUSY, both by F- and
D-terms, which opens the possibility of a dS vacuum thanks to the
$V_D$ contribution.  The F-auxiliary components of the $T$ and $M$
superfields are proportional to the K\"ahler derivatives. In this case,
$F_{T(M)}={\rm exp}(K/2)D_{T(M)}W$, with
\bea D_TW&=& W_T + K_T W = - {4\pi k_N\over N-N_f}W_{np} - {3\over 2
T_R}W \;, \nonumber\\ D_MW&=& W_M + K_M W = - {2 N_f M^{-1}\over
N-N_f}W_{np} +N_f \bar M\ W \;.
\label{Fterms}
\eea
The condition for unbroken SUSY by the previous F-terms is $D_T W =
D_M W=0$. One can immediately see that this requires
\be {\rm Unbroken\ SUSY}\ \Rightarrow \ |M|^2 = -{3\over 4\pi k_N
T_R}\;.
\label{SUSYcond}
\ee
Notice that, substituting eq.~(\ref{SUSYcond}) in $V_D$ [given by
eq.~(\ref{VDIIB})], and using the anomaly cancellation 
relation (\ref{anom3}), makes $V_D$ cancel.
This is a manifestation of the general connection (\ref{formula}) 
between F and D-terms. However, condition (\ref{SUSYcond})
{\em cannot} be fulfilled since the right hand side
is negative  definite.  Therefore, whichever the minimum of $V_F$ be,
it will correspond to a SUSY breaking point, with non vanishing $V_D$. 
This breaking of SUSY by the F-terms is in contrast with
the BKQ analysis, that did not consider the role of the matter fields
to implement gauge invariance. 
On the other hand, if we had added an extra 
singlet $\varphi$ to give mass to the
quarks, $V_D$ would have become immediately cancellable. To see this,
notice that the charge of $\varphi$ must have opposite sign to
$(q+\bar q)$ (and thus to $-\delta_{GS}$), so an appropriate VEV for
$\varphi$ is able to  cancel $V_D$. This confirms the results of
section~4 concerning the impossibility of having a non-cancellable $V_D$
if the quark fields have effective masses.

Let us now investigate in more detail the structure of the
scalar  potential. For the remainder of this analysis it is
particularly convenient  to split the complex fields $T$ and $M$ in
the following way
\bea T & = & T_R + {\rm i} T_I \nonumber \;, \\ M & = & |M| {\rm
e}^{{\rm i}  \alpha_M} \;.  \eea
Then eqs~(\ref{Fterms}) read 
\bea
D_T W  & = & -\frac{3 W_0}{2T_R} - \left(\frac{3(N-N_f)}{2T_R}
+4 \pi k_N\right) \left(\frac{2}{|M|^{2N_f}}\right)^{1\over N-N_f} 
{\rm e}^{-4 \pi k_N T_R\over N-N_f} 
{\rm e}^{{-2{\rm i}(N_f \alpha_M + 2\pi k_N T_I)
\over N-N_f}}  \;, \nonumber \\
&  \label{Fterms2} & \\
D_M W  &=& \frac{N_f}{M} \left[ |M|^2 W_0 + \left(|M|^2(N-N_f) - 2\right)
\left(\frac{2}{|M|^{2N_f}}\right)^{1 \over N-N_f}
{\rm e}^{-4 \pi k_N T_R\over N-N_f} 
{\rm e}^{{-2{\rm i}(N_f \alpha_M+2\pi k_N T_I)\over N-N_f}} 
 \right].   \nonumber
\eea
It is now straightforward to write $V_F$ plugging these expressions 
into eq.~(\ref{VFVD}). Due to the presence of the same phase,
$2(N_f \alpha_M+ 2\pi k_NT_I) /(N-N_f)$, in
$D_T W$ and $D_M W$, $V_F$ has the form
\be
\label{VF6}
V_F = A(T_R,|M|^2) + B(T_R,|M|^2) {\rm cos} 
\left[ \frac{2}{N-N_f} (N_f \alpha_M +  2\pi k_N T_I)\right] \;,
\ee
with 
\begin{eqnarray}
A(T_R, |M|^2) & = & \frac{{\rm e}^{N_f |M|^2}}{(2T_R)^3} 
\left[ N_f W_0^2 |M|^2 
+  \left(\frac{2}{|M|^{2N_f}}\right)^{2\over N-N_f} 
{\rm e}^{-8\pi k_N T_R\over N-N_f}  
\left(   \frac{(8 \pi k_N T_R)^2}{3}  \right. \right. 
\nonumber  \\
&&
\nonumber  \\
&+& \left. \left.  
16 \pi k_N T_R  (N-N_f) +  \frac{N_f}{|M|^2} [|M|^2(N-N_f)-2]^2  \right) \right]  
\\
&&
\nonumber  \\
 B(T_R,|M|^2) & = & \frac{{\rm e}^{N_f|M|^2}}{(2T_R)^3}  
\left(\frac{2}{|M|^{2N_f}}\right)^{1\over N-N_f} 
{\rm e}^{-4 \pi k_N T_R\over N-N_f}  \nonumber
\\
&&
\nonumber  \\
 & \times & 2 W_0 \left[ N_f |M|^2(N-N_f) 
-2N_f + 8 \pi k_N T_R  \right] \;\;.
 \end{eqnarray}
$V_F$ has extrema for
\be
\left[ \frac{2}{N-N_f} (N_f \alpha_M +  2\pi k_N T_I)\right]  = n \pi \;,
\label{phase}
\ee
with $n$ integer. Depending on the signs of $A$ and $B$, 
$n$ even (odd) will correspond to  a minimum (maximum) of $V_F$ 
or vice versa.
These are also extrema of the whole $V=V_F+V_D$ potential, since
$V_D$,
as given by eq.~(\ref{VDIIB}), does not depend on $\alpha_M, T_I$.
Therefore, one can integrate out one combination of these two real variables
through eq.~(\ref{phase}) (there is another combination which is 
completely flat) and work with $V(T_R,|M|^2)=V_F + V_D$, with 
$V_F$ and $V_D$ given by eq.~(\ref{VF6}) [with the cosine = --1,
 which corresponds to the solution of minimum in our cases] and eq.~(\ref{VDIIB}) respectively.

Next we explore the minimisation of this potential, searching for examples of dS vacua.

\subsection{Working examples. General characteristics}

Let us start by considering the previous set up
with the following choice of parameters: $N=15$, $k_N=1$, $N_f=1$, 
$q=\bar{q}=-2$, $W_0=0.30$. Incidentally, $N_f=1$ means that there 
is just one $M=\sqrt{2Q\bar Q}$ field,
so we are not making any overall squark condensate assumption.

The total  potential, $V_F+V_D$, as a function of the $T_R$ and $|M|^2$
variables (the other two, $\alpha_M, T_I$, have been integrated out as
described above) is plotted in figure~\ref{fd}, where one can see
the existence of a positive (dS) minimum, as desired.
\FIGURE{\centerline{
\includegraphics[width=8cm]{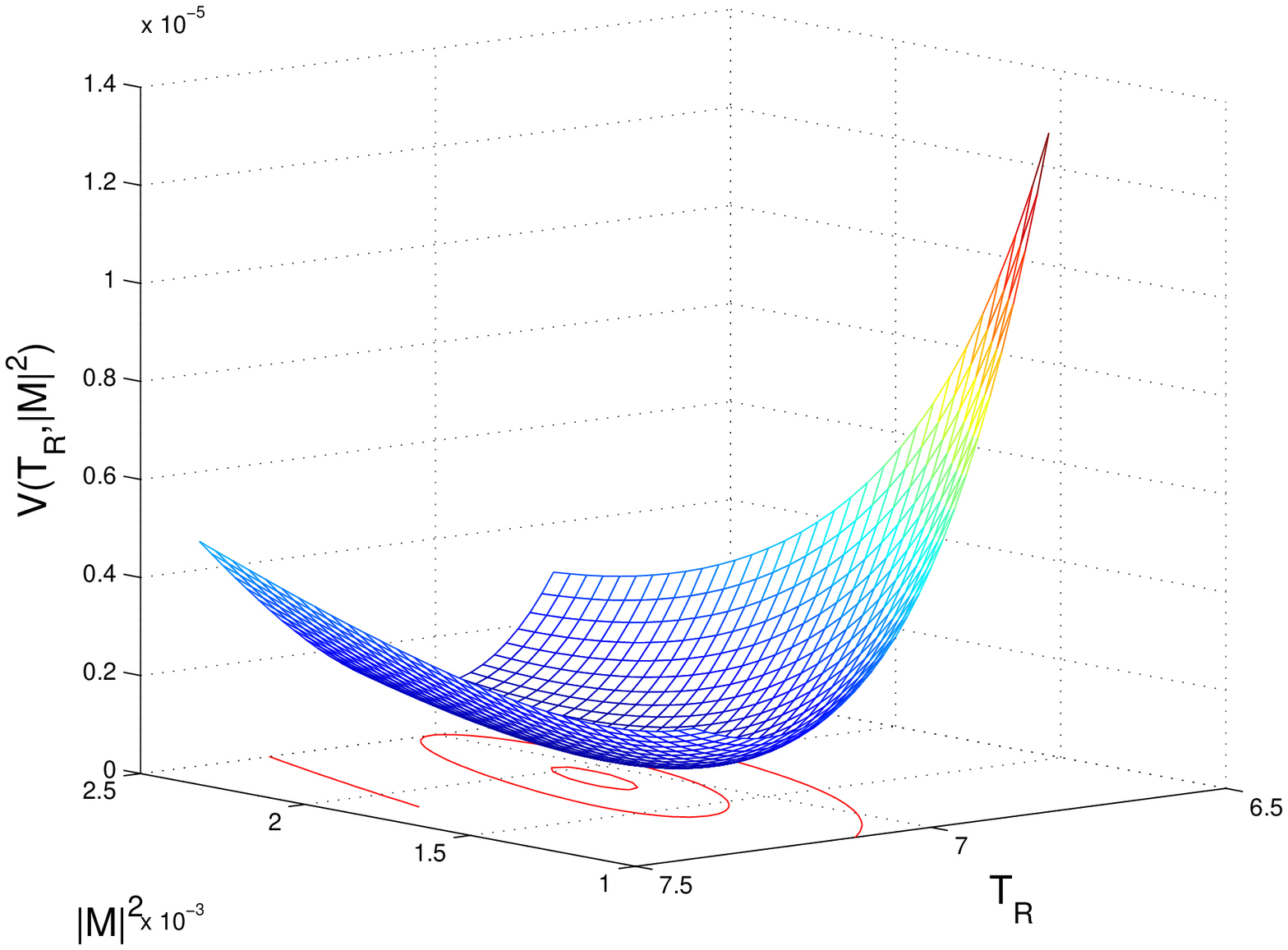}}
\caption{ 3D plot of the potential as a function of $T_R$ and $|M|^2$
for the values of the parameters shown in the text. Contour lines have
been projected on the base in order to see the minimum.}
\label{fd}}
A contour plot has been projected on the base in order to
appreciate better the position of the minimum, which happens for  $T_R
=7.07$, $|M|^2 = 0.00188$ and $V_F+V_D = 5.64.10^{-7}$.  We
have performed an analysis of the Hessian matrix in all four variables
to confirm that this is a real minimum of $V$.

It is perhaps more illustrative to show the contribution of each part
of the potential separately. In fig.~\ref{contf} we show a contour
plot  of $V_F$ in the $(T_R,|M|^2)$ plane.
\FIGURE{\centerline{ \includegraphics[width=8cm]{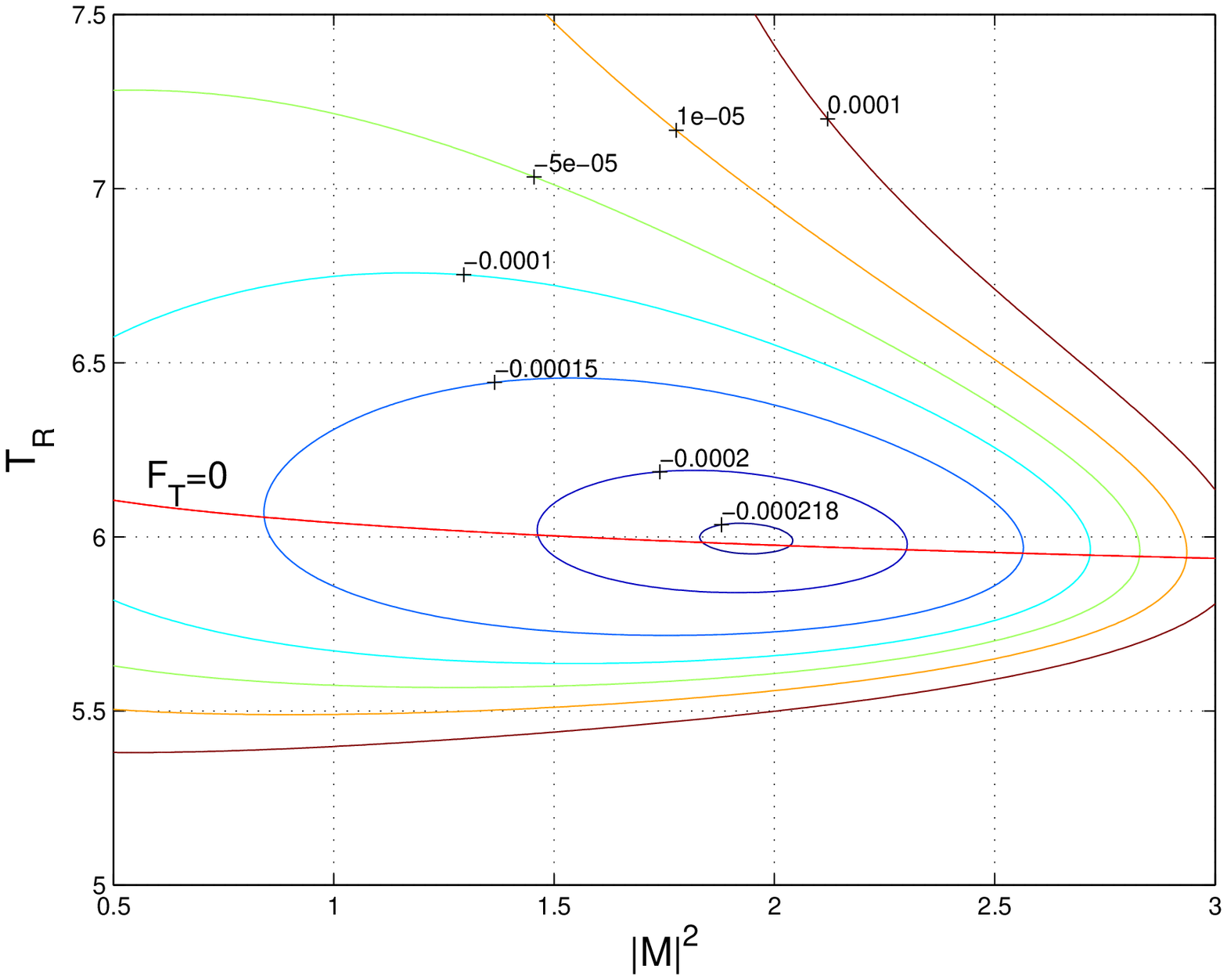}}
\caption{Contour plot of the F-potential, $V_F$,  as a function of $T_R$
and $|M|^2$ for the values of the parameters shown in the text. The
condition $F_T=0$ is also shown.}
\label{contf}} 
We can clearly see that this potential has a minimum for $T_R=5.99$,
$|M|^2 = 1.94$ and $V_F = -0.00022$. We have  also shown the
condition $F_T=0$ to illustrate the fact that the F-term associated to
the modulus does not break SUSY at this stage. However, the remaining F-term,
$F_M$, does break SUSY and,  in fact, the line $F_M=0$ lies outside
the boundaries of the figure. Therefore, the F-part of the potential
generates has a SUSY breaking minimum with negative vacuum energy, as
expected from the discussion of subsection~6.1.

As for $V_D$, this is shown in fig.~\ref{potd}.
\FIGURE{\centerline{ \includegraphics[width=8cm]{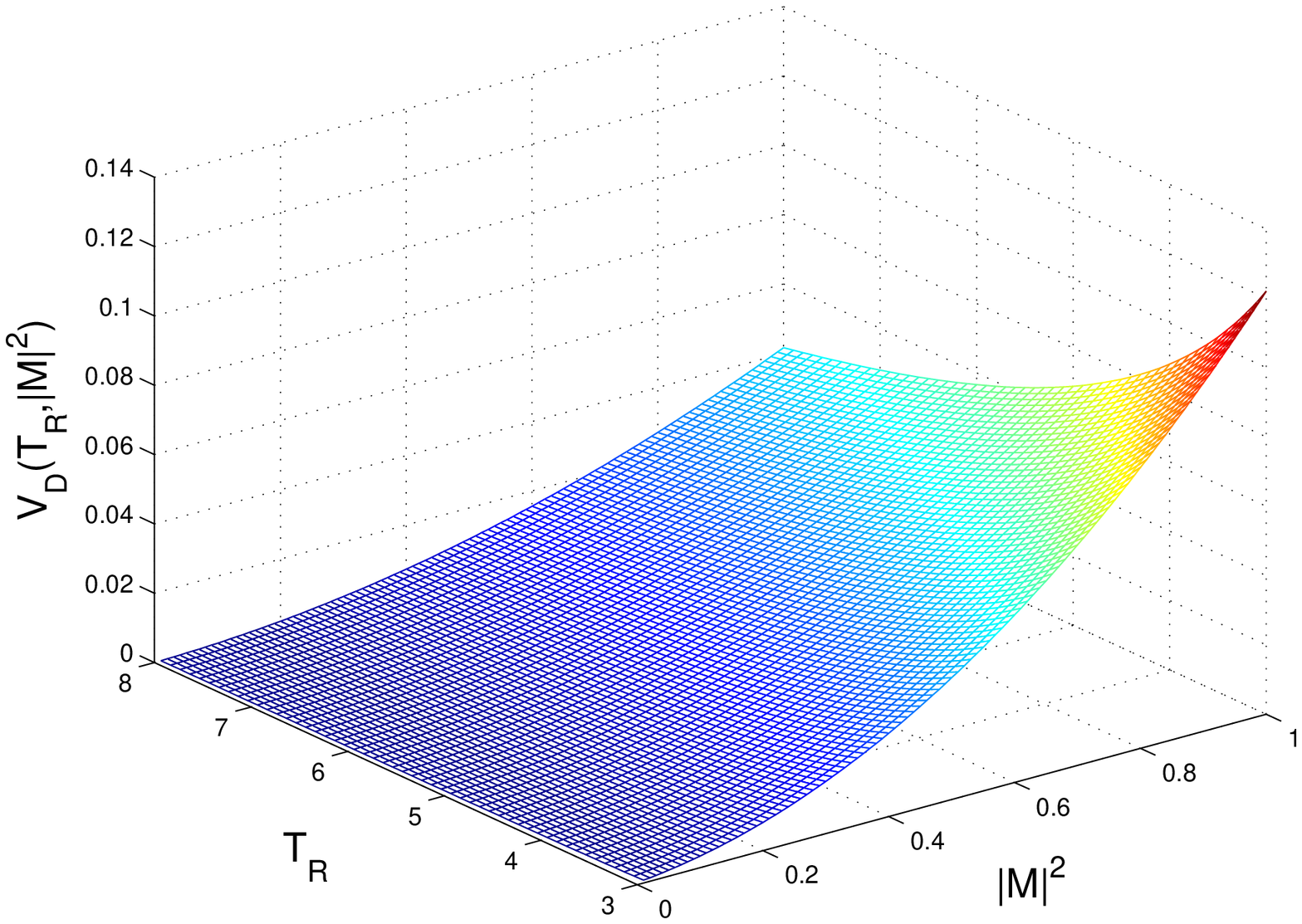}}
\caption{3D plot of the D-potential, $V_D$ as a function of $T_R$ and
$|M|^2$ for the values of the parameters shown in the text.}
\label{potd}}
Here we can see clearly that, along the $|M|^2$ direction, the minimum
lies at $|M|^2=0$, while there is a runaway direction along increasing
$T_R$. The sum of the two potentials is such that the minimum of
$V_F+V_D$ happens for similar values of $T_R$ as for $V_F$, but for much
smaller values of $|M|^2$.  The interplay between $V_F$ and $V_D$ to generate a dS minimum
can be well appreciated in fig.~4, where we have plotted $V_F$,
$V_D$ and $V=V_F+V_D$ as a function of  $T_R$ for $|M|^2$ fixed at its value in
the minimum of $V$.
\FIGURE{\centerline{ \includegraphics[width=8cm]{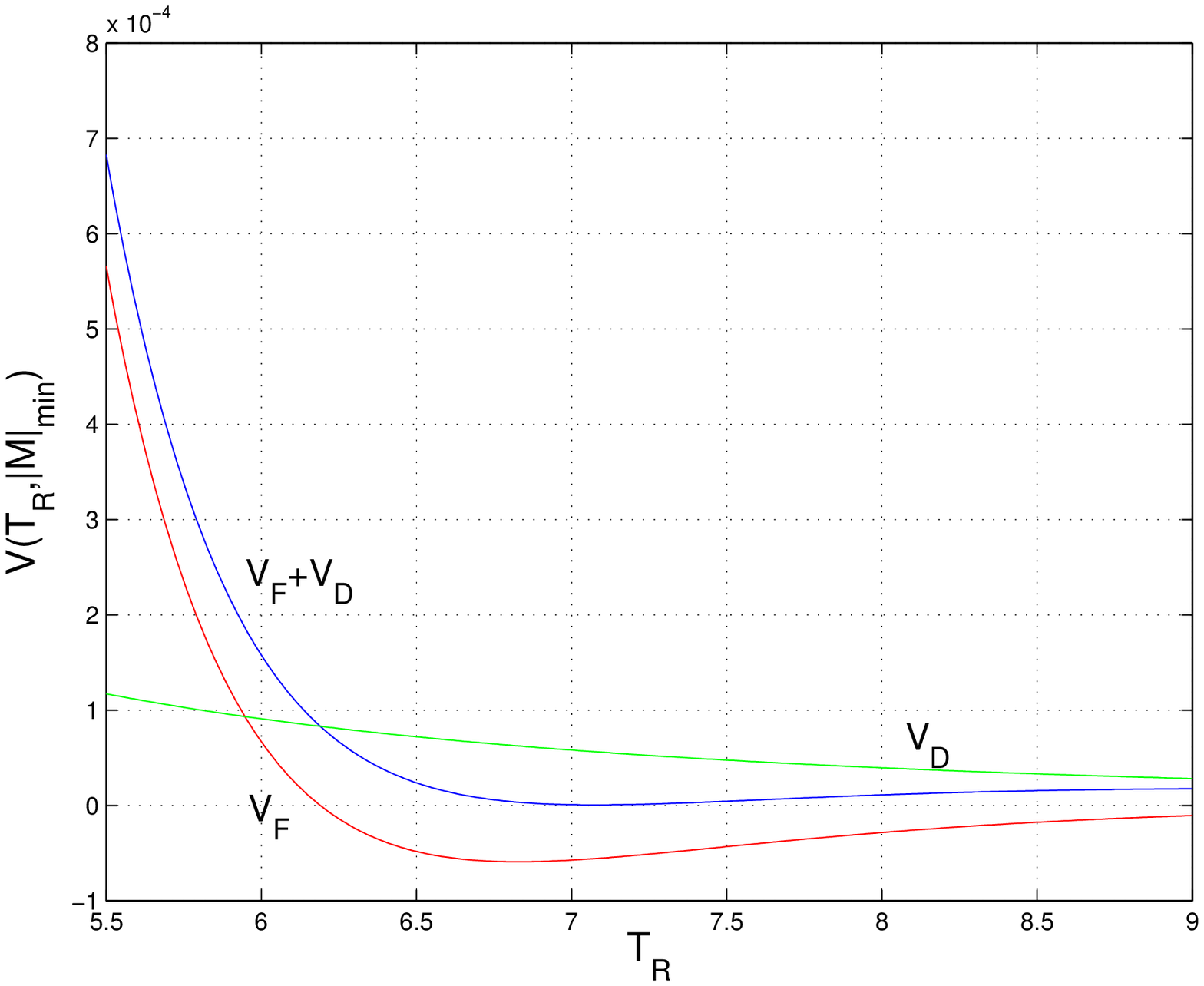}}
\caption{Plot of the F-potential, $V_F$ (red), the  D-potential, $V_D$ (green) and the sum of both, $V_F+V_D$ (blue), as a function of $T_R$ for the example shown in the text and $|M|^2$ fixed to its value at the overall minimum.}
\label{fig4b}}

We have found many more examples, with
reasonable values of the parameters, where this pattern is repeated. 
In table~\ref{tab:IIB} we present a handful of successful cases, 
including the one shown in the figures. The 'min' ('$F$min')
subscript means that the quantity is evaluated at the minimum
of $V$ ($V_F$).
\TABLE{
\begin{tabular}{|c|c|c|c|c|c|}
\hline
$N$ & --($q$, $\bar{q}$) & $W_0$ & $T_R|_{\rm min}$ ($T_R|_{F{\rm min}}$) & 
$|M|^2_{\rm min}$ ($|M|^2_{F{\rm min}}$) &  $V_F+V_D|_{\rm min}$  ($V_F|_{\rm min}$) \\ \hline \hline
15 & (2,2) & 0.30 & 7.07 (5.99) & 0.00188 (1.94) & 5.64.10$^{-7}$ (-0.000219) \\ \hline
14 & (1,2) & 0.25 & 6.83 (5.70) & 0.00169 (1.94) & 7.38.10$^{-6}$ (-0.000178) \\ \hline
13 & (2,3) & 0.27 & 6.29 (5.09) & 0.00161 (1.93)  & 2.04.10$^{-5}$ (-0.000286) \\ \hline 
12 & (3,3) & 0.28 & 5.92 (4.54) & 0.00129 (1.92) & 4.52.10$^{-5}$ (-0.000423) \\ \hline
18 & (3,4) & 0.26 & 8.97 (7.80) & 0.00134 (1.95) & 2.70.10$^{-7}$ (-7.82.10$^{-5}$)   \\ \hline 
13 & (3,-2) & 0.03 & 8.57 (7.48) & 0.00109 (1.95) & 1.34.10$^{-7}$ (-1.32.10$^{-6}$) \\ \hline
\end{tabular}
\caption{Viable examples with $N_f=1$, cos$=-1$ and $k_N=1$.}
\label{tab:IIB}}

A number of comments are in order:
 
\begin{itemize}

\item
The values found for $T_R$ are in the $5-10$ range 
(in $M_p$ units), which is 
satisfactory for the Supergravity approximation we are using to be
valid (which requires $T_R\simgt 1$), but also from the phenomenological
point of view. Notice here that the value of $T_R$ corresponds to
$g_{\rm YM}^{-2}(M_{\rm string})$ up to ${\cal O}$(1) 
model-dependent factors, called $k_a$ in sect.~5. If the
Standard Model (SM) gauge group arises from wrapped D7-branes (as 
we have assumed for the
condensing gauge groups) and all $k_a=1$, then one should get
$T_R\sim 12$. This is in fact not far from the values found
in the examples of table 1, where $k_N=1$ has been taken for
the condensing group, but no assumption about
the $k_a$ factors of the SM gauge groups
has been made. Therefore, playing with these $O(1)$ factors, it is
clear that the quoted values of $T_R$ are perfectly realistic.
On the other hand, as discussed in sect.~5, the SM gauge group
could alternatively arise from stacks of D3-branes
sitting at a point of the compactified space, in which case the
gauge coupling is given by the dilaton $S$, instead of $T$.
Then, the value of $T_R$ does not have a direct phenomenological
meaning and, again, the presented examples are perfectly viable.

\item The value of $W_0$ sets the overall scale of the F-potential,
and, consequently, determines the scale of the non-perturbative
potential, which has to be comparable in size in order to create a
minimum of $V_F$. Given that $W_0$ is an effective
parameter, coming from the flux stabilisation of the other moduli in
the model,  we treat it as a phenomenological quantity and
simply state the range of values for which we find suitable
minima, which, as can be seen from the table, is around
$W_0 \sim 0.3$ (for both anomalous charges having the same sign). 
Smaller values of $W_0$ result in a smaller
magnitude for $V_F$ which would imply that $V_D$ dominates and we lose
the minimum. Larger values would result in a too large $V_F$ and the
uplifting by $V_D$ would not be efficient enough. 

\item Playing 
with the value of $W_0$ one can find $V=0$ (i.e. Minkowski) vacua, 
or adjust the cosmological
constant to an arbitrary small value. E.g.
for the example shown in the figures we can tune $W_0=0.301566$ to
have a minimum at $T_R=7.07$, $|M|^2=0.0019$ and $V=1.10^{-9}$. The
value $W_0=0.301567$ would already render a negative
minimum. Roughly speaking each decimal place tuned in $W_0$ reduces
the value of $V$ at the minimum by half an order of magnitude.

\item The small values of $|M|^2$ allow, for the purpose of 
minimisation, to neglect $|M|^2$
in $V_D$. In this
way one exactly recovers the BKQ potential (\ref{VDFernando2}).
However, this does not mean that one can simply replace $|M|^2$
by a constant in $W_{np}$ and proceed as BKQ. The reason is
that the contribution of the $M$-field to the $V_F$ potential, 
$e^K |D_M W|^2\sim k/T_R$ (where $k$ is a constant), is sizeable 
and cannot be ignored in the minimisation. Likewise, the 
dependence of $V$ in both fields, $T$ and $M$, may be relevant
to inflation.

\item To obtain $T_R > 1$ at the minimum, a value of $N \geq 10$ has
to be assumed. This is due to the fact that the
minimum for the F-potential comes from the interplay between $W_0$
and $W_{np}$, which 
determines the order of magnitude of the exponent of $W_{np}$, 
$4\pi k_N/(N-1)$,
in terms of $W_0$ and $T_R$. Then, for the quoted size
of $W_0$ and $T_R>5$, $N$ must be at least of order $12$, as shown.

\item The values of the anomalous charges ($q$, $\bar{q}$) affect
the scale of the D-potential and must, therefore, be balanced 
with the corresponding value of $W_0$ for the uplifting 
mechanism to be efficient.

\item We have found $F_T\sim0$ at the minimum of $V_F$ in all examples.
This leads to a simple expression for $|M|^2_{F{\rm min}}$, given by
\be
\left( \frac{2}{|M|^2_{F{\rm min}}} \right)^{1/(N-1)} 
= \frac{\frac{3W_0}{2T_R}}{\frac{3(N-1)}{2T_R}+4 \pi k_N} 
{\rm e}^{4 \pi k_NT_R/(N-1)}  \;\;.
\ee
For the values of parameters considered one can check that 
the previous equation  
leads to a value of 
$|M|^2_{F{\rm min}} \sim 2$ which is remarkably stable.
As commented before, this value gets very suppressed once $V_D$ is added.

\item 
After adding the $V_D$ piece, both $D_TW$ and  $D_MW\ \neq\ 0$.
Hence, SUSY becomes broken by the $T$ and $M$ F-terms, and also
by the D-term associated to $U(1)_X$.
The goldstino field can be written as $\eta=c_M \tilde M
+ c_T\tilde T + c_\lambda \lambda_X$, where $\tilde T$, $\tilde M$ are 
the (canonically normalised) fermionic components of $T$, $M$,
and $\lambda_X$ is the (canonically normalised) gaugino associated
to $U(1)_X$. For the example shown in the figures, the coefficients
$\{|c_M|, |c_T|, |c_\lambda|\}$ are in the ratio $\{1:25:42\}$. The 
resulting gravitino mass is naturally too large for ordinary 
low-energy SUSY phenomenology. E.g. for the same example as above, 
$m_{3/2} = {\rm exp}(K/2)W\simeq M_p/177$. This large $m_{3/2}$
is due to the magnitude of $V_D\sim [{\cal O}(M_p^2)/(4\pi^2)]^2$,
which has to be comparable
to the absolute size of $V_F$ in order to uplift the AdS minimum,
without destroying it. This may be a phenomenological shortcoming if
one attempts to get conventional low-energy SUSY. We leave
some further comments on this issue for the conclusions.

\end{itemize}

\noindent
In summary, playing with a constant superpotential (triggered by fluxes),
a non-renormalizable superpotential
originated by gaugino condensation, with quark
representations, and an anomalous $U(1)$ group with 
non-cancellable FI D-term (which, for 
gauge invariance consistency, requires the quarks to be massless),
it is easy to find examples where an AdS minimum is produced by 
the F-potential, and subsequently uplifted to dS by the D-term.
There is no fine tuning in our
results (unless one wishes to fine-tune the cosmological constant) 
and, in practice, this fixes the acceptable ranges for $W_0$ 
and the rank of the gauge group.

\section{Conclusions}
 \label{sec:conclu}

In this article we have revisited the proposal of BKQ~\cite{Burgess:2003ic} of 
using a supersymmetric D-term potential
(namely a Fayet-Iliopoulos one) for the uplifting of the AdS minima
found by KKLT~\cite{Kachru:2003aw} in the context of type IIB
theory with fluxes. The BKQ suggestion has been criticised
\cite{Choi:2005ge,deAlwis:2005tf} since, on general SUGRA
grounds, a model with vanishing F-terms must have also 
vanishing D-terms, which prevents the D-part of the potential ($V_D$)
from uplifting the SUSY-preserving minima of the F-part ($V_F$).

First, we have reconciled the BKQ scenario with the general SUGRA
arguments, by making the former gauge invariant. This requires
the inclusion of matter fields, which play a crucial role for
the consistency of the approach. In this context we show that
a non-perturbative superpotential $\sim e^{-aT}$ (as required
by the KKLT set up) produced by gaugino condensation is only
consistent with a non-cancellable $V_D$ (as required by the 
BKQ proposal) if the relevant quark representations are massless.
Then the minima of $V_F$ are necessarily SUSY-breaking, either 
by the moduli F-terms or by the matter ones, and the uplifting by
$V_D$ can in principle work. We discussed also the details of such
effective SUGRA scenarios when they arise from type IIB or from
the heterotic strings, paying special attention to the anomaly
cancellation constraints.

Finally, we illustrated and applied the previous results,
by finding many examples of effective SUGRA models (which can arise
from type IIB strings in the KKLT context), whose potential has
positive minima for reasonable values of the moduli and matter
fields. This shows how the uplifting by D-terms works in
practice. Before adding the $V_D$ piece to the potential, these
examples have SUSY broken by the F-term associated to the matter
fields, whereas $F_T\sim0$. The minima are initially AdS and become dS
after adding the D-terms. The corresponding values of the parameters
that define the model lie within natural ranges and are not fine tuned
(unless one wishes a vanishing or extremely tiny cosmological
constant, which is perfectly achievable).

At the minimum of the complete potential the breaking of SUSY is 
triggered both by the F-terms (i.e. $F_M,\ F_T\neq\ 0$) and by 
the anomalous D-term ($D_X\neq 0$).
The gravitino mass is ${\cal O}(10^{-2}M_p)$, since the natural
scale of $V_D$ (and hence of $V_F$ at the uplifted minimum) is 
$[{\cal O}(M_p/(4\pi^2))]^2$. 
This large gravitino mass is an obstacle for conventional
low-energy SUSY (although it is OK for inflation), since it leads to
${\cal O}(m_{3/2})$ soft masses in the observable sector, after
gravity mediation of SUSY breaking. A possibility here is simply to
give up low energy SUSY.
However, a particularly appealing possibility arises from
noting that, although the scalar soft masses are naturally very large
in this framework, the (observable) gaugino masses are vanishing at
this stage if the SM gauge group arises from stacks of D3-branes
sitting at a point of the compactified space 
(since $T$ does not enter the relevant gauge kinetic functions).
It is amusing that this scenario (which, of course, would get radiative
corrections) corresponds to the assumptions of the so-called
split-SUSY models \cite{Arkani-Hamed:2004fb,Arkani-Hamed:2004yi}, 
whose possible existence and viability
has been invoked precisely in the context of landscape frameworks, as
those suggested by the KKLT set up re-visited here.

\acknowledgments
                                                                               
We thank A. Font, P. Garc\'{\i}a del Moral, L. Ib\'a\~nez, O. Seto and
A. Uranga for extremely useful discussions and advice. AA is supported
in part by the Netherlands Organisation for Scientific Research
(N.W.O.) under the VICI Programme, and by the Spanish Ministry of
Education through projects FPA2002-02037 and FPA2005-04823. The work
of BdC is supported by PPARC, and that of LD by Fondazione Angelo
Della Riccia. JAC thanks the Physics and Astronomy department at
Sussex University for hospitality and The Royal Society for support
through a Visiting Fellowship.

\end{document}